\documentclass[twocolumn,aps,prl]{revtex4-2}
\usepackage{graphicx}
\newcommand*{\NwarrowA}{\rotatebox[origin=c]{135}{\(\nLeftrightarrow\)}}
\newcommand*{\NwarrowB}{\rotatebox[origin=c]{45}{\(\nLeftrightarrow\)}}
\newcommand*{\nUpdownarrow}{\rotatebox[origin=c]{90}{\(\nLeftrightarrow\)}}
\newcommand{\highlight}[1]{\colorbox{yellow}{$\displaystyle #1$}}

\usepackage{amsmath}
\usepackage{url}
\usepackage[english]{babel}
\usepackage{chngcntr}
\usepackage{natbib}
\usepackage{textcase}
\usepackage{hyperref}
\usepackage{tikz}
\usepackage{xcolor}
\usepackage{epstopdf}
\usepackage[percent]{overpic}
\usepackage{cancel}
\usepackage{multirow}
\usepackage{amssymb}
\hypersetup{
    colorlinks,
    citecolor=blue,
    filecolor=blue,
    linkcolor=blue,
    urlcolor=blue
}

\begin{document}

\title{Entangling spin and charge degrees of freedom in semiconductor quantum dots}

\author{Marko J. Ran\v{c}i\'{c}}
\email{marko.rancic@totalenergies.com}
\affiliation{TotalEnergies, Nano-INNOV – Bât. 861, 8, Boulevard Thomas Gobert – 91120 Palaiseau – Francee}
\date{\today}

\begin{abstract}
In this theoretical manuscript I propose a scheme for entangling a single electron semiconductor spin qubit with a single electron semiconductor charge qubit in a triangular triple quantum dot configuration. 
Two out of three quantum dots are used to define a single electron semiconductor charge qubit. Furthermore, the spin qubit is embedded in the Zeeman sub-levels of the third quantum dot.
Combining single qubit gates with entangling CNOT gates allows one to construct a SWAP gate, and therefore to use the semiconductor spin qubit as a long-lived memory for the semiconductor charge qubit.
\end{abstract}
\maketitle

\section{1) Introduction} Since it was concluded that bits obeying the laws of quantum mechanics offer advantages over bits obeying the laws of classical physics, many systems have emerged as possible candidates for quantum mechanical bits (qubits) \cite{Cirac1,Loss1,Kane1,Nakamura1,Divincenzo1,Vandersypen1,Levy1,Hayashi1,Chiorescu1,DasSarma1}.
Semiconductor materials like GaAs and Si, used to mass produce electronic components for over half a century, also show prospect for embedding a qubit. There are two common ways in which a single electron qubit can be embedded in a semiconductor quantum dot, 
using the spin \cite{Loss1} or the charge \cite{Hayashi1} degree of freedom of the electron. 

Charge occupancy of a double quantum dot can be used to define a single electron semiconductor charge qubit \cite{Hayashi1,Gorman1,Shinkai1,Petersson1,Dovzhenko1,Cao1,Shi1,Kim1}. Although the semiconductor charge qubit can be controlled on fast timescales \cite{Petersson1}
(compared to many other implementations of a qubit), it
suffers from a relatively short coherence time, due to the interactions with its noisy, semiconductor environment \cite{Dovzhenko1}.
Furthermore, the single electron spin qubit can be defined in Zeeman sub levels of the electron spin \cite{Loss1}. And although the coherence times of a single electron spin qubits (Loss-DiVincenzo qubit) have been gradually increasing up to hundreds of microseconds \cite{Petta1, Bluhm1, Veldhorst1}, 
single spin qubit operations are much slower compared to those of a charge qubit [\onlinecite{Koppens1}, \onlinecite{Yoneda1}].

Creating entanglement between disparate degrees of freedom \cite{Sorensen1, Zhu1, Fuchs1, Gao1, deGreve1, PhysRevB.94.045316, PhysRevB.99.165306} is a crucial ingredient in creating long-lived memories \cite{Simon1}, quantum repeater networks \cite{Briegel1} and quantum teleportation [\onlinecite{Bennet1}, \onlinecite{Bouwmeester1}]. 
In this manuscript I present an entangling scheme for a single electron charge degree of freedom with a single electron spin degree of freedom. The entanglement is achieved via a control NOT (CNOT), giving rise to
a maximally entangled Bell state $|\Psi^-\rangle=(|0\rangle_c |0\rangle_s-|1\rangle_c |1\rangle_s)/\sqrt{2}$ of the spin $s$ and charge $c$ degrees of freedom.
Combining three entangling CNOT gates with single qubit Hadamard gates allows us to construct a SWAP gate [\onlinecite{Nielsen1}]. Furthermore, a SWAP gate allows us to encode a general superposition of single electron charge states 
of a Double Quantum Dot (DQD) into the spin degree of freedom of an electron residing in a nearby quantum dot. This could allow coupling between distant semiconductor charge qubits (mediated by semiconductor spin qubits \cite{deGreve1, PhysRevX.2.011006, mi2018coherent}), or allow for full scale quantum computation based on semiconductor charge qubits with a spin qubit being used as a long lived memory. This approach exploits the best properties of spin and charge qubits, the long coherence times of spin qubits and the fast control of the charge qubit. 
By embedding the information about the charge qubit into the spin degree of freedom, the state of the charge qubit survives for up to a thousand charge coherence times. 

\begin{figure}[t!]
\centering
\includegraphics[width=0.3\textwidth]{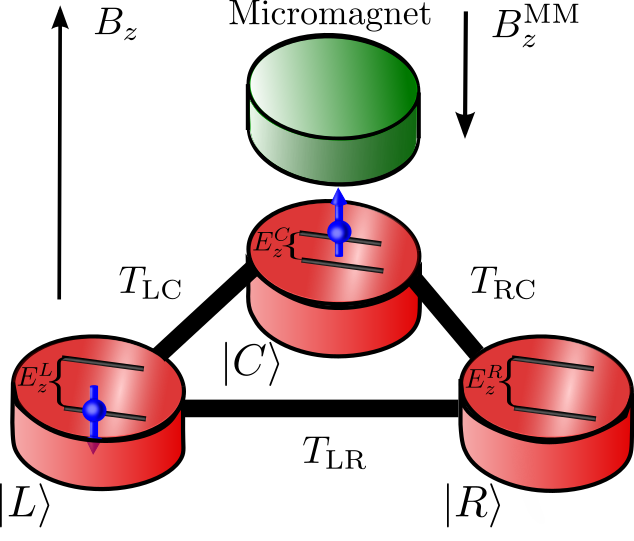}
\caption{(Color online) The setup comprises of a triangular triple quantum dot in an external magnetic field $B_z$ loaded with two electrons. One of the quantum dots has a micromagnet embedded on top $B_z^{\rm MM}$. The tunnel barriers between the dots (depicted with black lines, with their amplitudes denoted with $T_{\rm LC}$, $T_{\rm CR}$, $T_{\rm LR}$) need not be controlled independently.}
\label{Setup}
\end{figure}

\section{2) Setup} The setup consists of two electrons in a triangular triple quantum dot in an external magnetic field, with a micromagnet embedded on top of one of the quantum dots Fig. \ref{Setup}. 
The spin qubit is defined in Zeeman sublevels of the quantum dot with the micromagnet on top ${|0\rangle_s \longrightarrow |C,\uparrow\rangle}$, ${|1\rangle_s \longrightarrow |C,\downarrow\rangle}$. From now on I will refer to the spin qubit quantum dot 
as the central dot. Furthermore, the logical subspace of the charge qubit is defined as the charge occupancy of the remaining two dots, with the electron spin in the spin down state 
${|0\rangle_c \longrightarrow |L,\downarrow\rangle}$ and ${|1\rangle_c \longrightarrow |R,\downarrow\rangle}$. From now on I will refer to the charge qubit quantum dots as the left and right dot.

\begin{figure}[t!]
	\centering
	\includegraphics[width=0.4\textwidth]{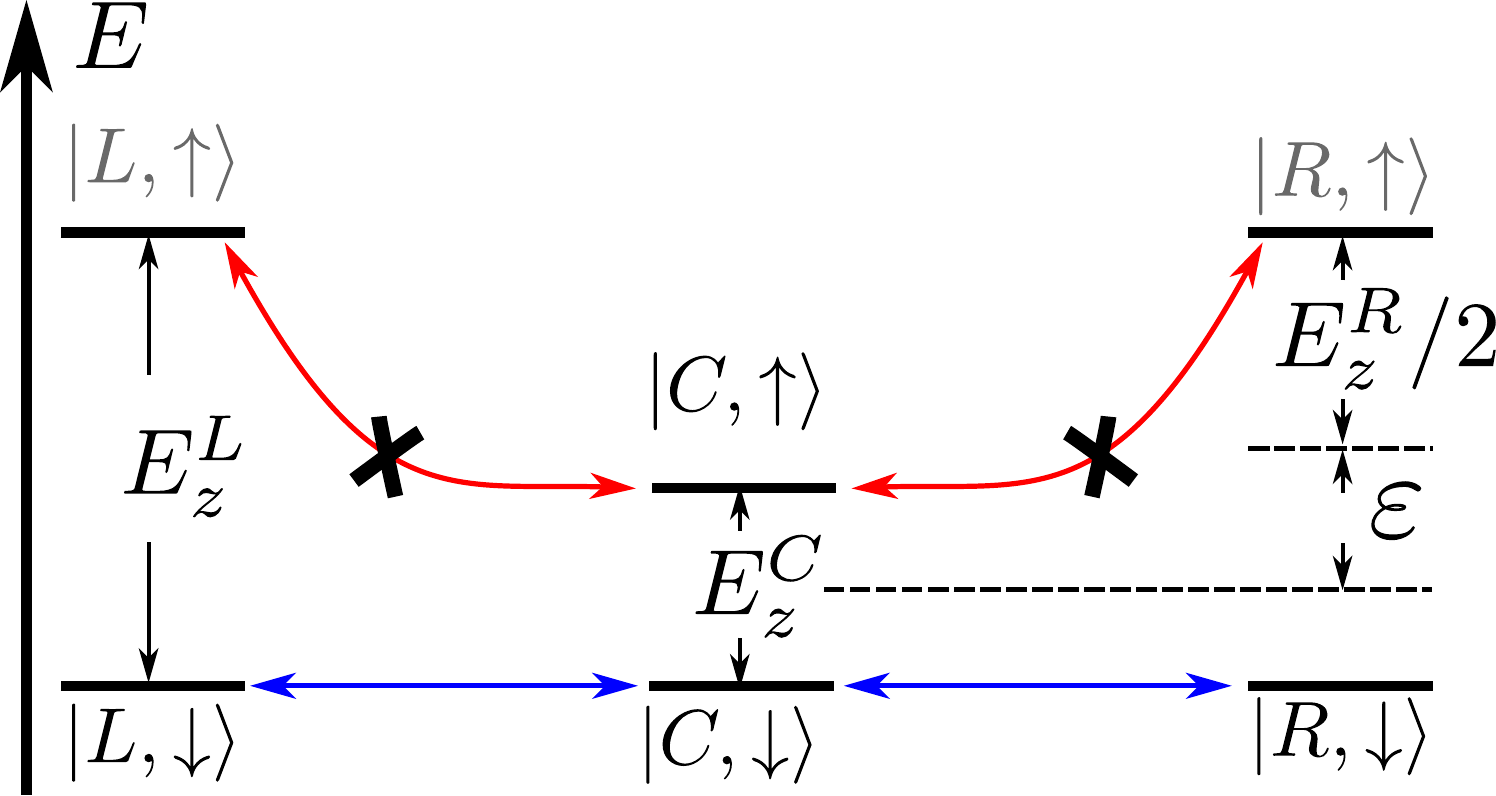}
	\caption{(Color online) The energy diagram, the charge qubit states $\{|L,\downarrow\rangle,|R,\downarrow\rangle\}$, the spin qubit states $\{|C,\downarrow\rangle,|C\uparrow\rangle\}$. 
	The center quantum dot (spin qubit quantum dot) is biased ${\varepsilon_C=-(E_z^{L,R}-E_z^{C})/2}$.}
	\label{EnDig}
\end{figure}

I assume that the tunnel coupling between the dots can be selectively controlled. 
The maximally entangled Bell state $|\Psi^-\rangle={(|0\rangle_c |0\rangle_s-|1\rangle_c |1\rangle_s)/\sqrt{2}}$ of spin and charge degree of freedom is achieved by initializing a ${|L,\downarrow;C,\downarrow\rangle}$, followed by
biasing the center quantum dot, see Fig \ref{EnDig}. Furthermore, a single spin Hadamard operation is applied, followed by tunneling events between the L and the C dot and the C and the R dot marked with tunnel hoppings $T_{LC}$, $T_{CR}$ respectively.

The central dot is biased for $\varepsilon_C$, 
so that the states $|L,\downarrow\rangle$, $|C,\downarrow\rangle$ and $|R,\downarrow\rangle$ are at the same energy, 
${E_{L,\downarrow}=E_{C,\downarrow}=E_{R,\downarrow}}$ Fig. \ref{EnDig}. Furthermore, the transitions ${|L,\uparrow\rangle\longrightarrow|C,\uparrow\rangle}$ and ${|C,\uparrow\rangle\longrightarrow|R,\uparrow\rangle}$ 
are energetically not favored when ${|E_{L,\uparrow}-E_{C,\uparrow}|=|E_{C,\uparrow}-E_{R,\uparrow}|\gg T_{ij}}$.

The shuttling time $\tau$ is chosen so that the initial electronic state ${|L,\downarrow;C,\downarrow\rangle}$ gets transferred to ${|C,\downarrow;R,\downarrow\rangle}$ with unit probability. In order to calculate the shuttling time $\tau$ I constrain myself to the 3-dimensional subspace, in which all electrons are in the spin down state. The effective three dimensional Hamiltonian describing such a process is defined \cite{Renzoni1, Brune1, Brune2}

\begin{multline}
H^{3{\rm D}}=\sum\limits_{i}(\varepsilon_i+E^i_{z\downarrow})n_{i\downarrow}+V\sum\limits_{\langle ij \rangle}n_in_j\\
+(T_{LC}c_{L\downarrow}^{\dagger}c_{C\downarrow}+T_{CR}c_{C\downarrow}^{\dagger}c_{R\downarrow}+H.c.).
\end{multline}
Here, $i,j=\{L,C,R\}$, $n_i=c_{i\downarrow}^{\dagger}c_{i\downarrow}$, and $c_{i\downarrow}^{\dagger}$ and $c_{i\downarrow}$ are the electronic creation and 
annihilation operators. Furthermore, $E_z^i$ is the Zeeman energy of the $i$th dot and $\varepsilon_i$ is the static bias. In the remainder of this paper I assume for simplicity that $T_{LC}=0$ or $T$ and $T_{CR}=0$ or $T$.

After inserting the time-independent Hamiltonian into the time dependent Schr\"{o}dinger equation I obtain the following set of coupled equations for the probability amplitudes
\begin{center}
\begin{equation}\label{eq:AmpEq}
\dot{a}_{LC}=i\frac{T}{\hbar}a_{LR},\,\dot{a}_{CR}=i\frac{T}{\hbar}a_{LR},\,\dot{a}_{LR}=i\frac{T}{\hbar}(a_{LC}+a_{CR}).
\end{equation}
\end{center}
After solving the system Eq. (\ref{eq:AmpEq}) for initial probability amplitudes $a_{LC}(0)=1$, $a_{CR}(0)=0$, $a_{LR}(0)=0$, I obtain the following occupation probabilities, with $P_{ij}(t)=|a_{ij}(t)|^2$
\begin{flushleft}
\begin{equation*}
P_{LC}=\frac{1}{4}\left(1+\cos{\left(\frac{T t \sqrt{2}}{\hbar}\right)}\right)^2,\,P_{LR}=\frac{1}{2}\sin^2{\left(\frac{T t \sqrt{2}}{\hbar}\right)},
\end{equation*}
\begin{equation}\label{eq:ProbEq}
P_{CR}=\frac{1}{4}\left(1-\cos{\left(\frac{T t\sqrt{2}}{\hbar}\right)}\right)^2.
\end{equation}
\end{flushleft}

From Eq. (\ref{eq:ProbEq}) we see that the probability that the electron transfers from $|L,\downarrow;C,\downarrow\rangle$ to $|C,\downarrow;R,\downarrow\rangle$ 
($P_{LC}(t)=0$, $P_{CR}(t)=1$, $P_{LR}(t)=0$) is satisfied for $t=\pi\hbar/T\sqrt{2}$, and I refer to this time as $\tau$ in the remainder of the manuscript. 

Now we will proceed to a derivation of the effective CNOT gate between the spin and charge degrees of freedom. The starting point in describing the system more quantitatively is the Hubbard Hamiltonian in the second quantization which is given by
\begin{multline}\label{eq:Ham0}
H_0=\sum\limits_{i\sigma}(\varepsilon_i+E_z^i{\bf \sigma})n_{i\sigma}+U\sum\limits_i n_{i\uparrow} n_{i\downarrow}+V\sum\limits_{\langle ij \rangle}n_in_j,
\end{multline}
where, $E_z^i$ is the Zeeman energy of the $i$-th quantum dot, $\varepsilon_i$ the bias of the $i$-th dot and $n_i=n_{i\uparrow}+n_{i\downarrow}=c_{i\uparrow}^{\dagger}c_{i\uparrow}+c_{i\downarrow}^{\dagger}c_{i\downarrow}$. Much like in Fig. \ref{EnDig} I set $\varepsilon_L=\varepsilon_R=0$ and ${\varepsilon_C=-(E_z^{L,R}-E_z^{C})/2}$.

The electrons can tunnel between the dots described by the following Hamiltonian \cite{Renzoni1, Brune1, Brune2}
\begin{equation}\label{eq:HamInt}
H_{ij}= T\sum \limits_{\langle 
	ij\rangle\sigma} c_{i\sigma}^{\dagger}c_{j\sigma}.
\end{equation}
In Eq. (\ref{eq:Ham0}) and Eq. (\ref{eq:HamInt}) $i,j={\{|L\rangle,|C\rangle,|R\rangle\}}$, $\sigma={\{|\!\downarrow\rangle,|\!\uparrow\rangle\}}$. Enabling only tunneling between left-to-center and center-to-right dots we obtain the following Hamiltonian
\begin{equation}\label{eq:hamTot}
\tilde{H}=H_0+T (c^{\dagger}_{L\sigma} c_{C\sigma}+c^{\dagger}_{ C\sigma} c_{R\sigma}+{\rm H.c.}).
\end{equation}
Given that $U\gg V$ \cite{PhysRevB.59.2070} we set $V$ to zero in this toy derivation. The $m_s=-1$, $m_s=+1$ and $m_s=0$ blocks are decoupled in the absence of operations on the spin qubit. The time evolution operator takes the form $\mathcal{U}=\exp{\left(-i\tilde{H}t/\hbar\right)}$. In our study the operator $\tilde{H}$ has a block diagonal structure leading to the following formula

\begin{table}[h!]
	\begin{tabular}{c|c|ccccc}
		
		$L\uparrow C\uparrow$&     $\highlight{L\downarrow C\downarrow}$&  $L\uparrow C\downarrow$&  \textcolor{orange}{$\nLeftrightarrow$}&  $L\uparrow L\downarrow$&  \textcolor{orange}{$\nLeftrightarrow$}&  $\highlight{L\downarrow C\uparrow}$\\
		\textcolor{red}{$\nUpdownarrow$} & $\Updownarrow$&$\Updownarrow$ & \textcolor{orange}{$\NwarrowA$}& & \textcolor{orange}{$\NwarrowB$}&  \textcolor{red}{$\nUpdownarrow$}\\
		$L\uparrow R\uparrow$&     $L\downarrow R\downarrow$&  $L\uparrow R\downarrow$&  &  $C\uparrow C\downarrow$&  &  $L\downarrow R\uparrow$\\
		\textcolor{red}{$\nUpdownarrow$}& $\Updownarrow$&\textcolor{red}{$\nUpdownarrow$} & \textcolor{orange}{$\NwarrowB$}& & \textcolor{orange}{$\NwarrowA$}&  $\Updownarrow$\\
		$C\uparrow R\uparrow$&     $\highlight{C\downarrow R\downarrow}$&  $\highlight{C\uparrow R\downarrow}$& \textcolor{orange}{$\nLeftrightarrow$} &  $R\uparrow R\downarrow$& \textcolor{orange}{$\nLeftrightarrow$} &  $C\downarrow R\uparrow$\\
		\hline
		$m_s=+1$& $m_s=-1$& & & $m_s=0$& &
		
	\end{tabular}
	\caption{The scheme of couplings of the system. States of the qubit are denoted in yellow. Couplings denoted by a canceled red double arrow are suppressed due to $|E_{C,\downarrow}-E_{R,\downarrow}|\gg T_{ij}$. Couplings denoted by a canceled orange double arrow are suppressed due to $U\gg T_{ij}$. Couplings denoted by a black arrow are resonant. $m_s$ denotes the spin projection quantum number. 
		\label{tab:TabX}}
\end{table}

\begin{equation}
\mathcal{U}=
\begin{pmatrix}
e^{-i\frac{t}{\hbar}\tilde{H}_{m_s=-1}}&& 0 &&  0\\
0 &&e^{-i\frac{t}{\hbar}\tilde{H}_{m_s=1}} && 0\\
0 && 0 && e^{-i\frac{t}{\hbar}\tilde{H}_{m_s=0}}
\end{pmatrix}.
\end{equation}When ${\varepsilon_C=-(E_z^{L,R}-E_z^{C})/2}$, ${|E_{C,\uparrow}-E_{R,\uparrow}|\gg T_{ij}}$ ${U\gg T_{ij}}$ the $m_s=+1$ block of the time evolution operator is a diagonal unit matrix $\mathcal{U}_{m_s=+1}=\text{diag}(1,1,1,1)$. 

In the same parameter regime and when ${T_{LC}=T_{CR}=T}$ the $m_s=-1$ block of the time evolution operator is
\begin{widetext}
\begin{equation}
\mathcal{U}_{m_s=-1}=
\begin{pmatrix}
	\frac{1}{2}+\frac{1}{2}\cos{\left(\frac{Tt\sqrt{2}}{\hbar}\right)}&& \frac{i}{\sqrt{2}} \sin{\left(\frac{Tt\sqrt{2}}{\hbar}\right)}&&  -\frac{1}{2}+\frac{1}{2}\cos{\left(\frac{Tt\sqrt{2}}{\hbar}\right)}\\
	\frac{i}{\sqrt{2}} \sin{\left(\frac{T_t\sqrt{2}}{\hbar}\right)} &&\cos{\left(\frac{Tt\sqrt{2}}{\hbar}\right)}&& \frac{i}{\sqrt{2}} \sin{\left(\frac{Tt\sqrt{2}}{\hbar}\right)}\\
	-\frac{1}{2}+\frac{1}{2}\cos{\left(\frac{Tt\sqrt{2}}{\hbar}\right)} && \frac{i}{\sqrt{2}} \sin{\left(\frac{Tt\sqrt{2}}{\hbar}\right)} && \frac{1}{2}+\frac{1}{2}\cos{\left(\frac{Tt\sqrt{2}}{\hbar}\right)}
\end{pmatrix},
\end{equation}
\end{widetext}
in the ${\{|L,\downarrow;C,\downarrow\rangle,|L,\downarrow ;R,\downarrow\rangle,\,|C,\downarrow; R,\downarrow\rangle\}}$ basis. When $t=\tau=\pi\hbar/T\sqrt{2}$ this becomes

\begin{equation}
\mathcal{U}_{m_s=-1}=
\begin{pmatrix}
0 && 0 &&  -1\\
0 &&-1 && 0\\
-1 && 0 && 0
\end{pmatrix}.
\end{equation}
The $m_s=0$ block has 9 states in total (Tab. \ref{tab:TabX}), so an analytical formula for the time evolution operator cannot be obtained. When $U \gg T$ the ${L\uparrow C\downarrow , \, L\uparrow R\downarrow , \, C\uparrow R\downarrow}$ (left column of the $m_s=0$ states in Tab. \ref{tab:TabX}) is decoupled from the doubly occupied states (middle column of $m_s=0$ states in Tab. \ref{tab:TabX}) which is in turn decoupled from the ${L\downarrow C\uparrow , \, L\downarrow R\uparrow , \, C\downarrow R\uparrow}$ states. This decoupling of single and double occupied states can effectively be described by setting the tunneling between those states to zero. The time evolution operator of the blocks containing ${L\uparrow C\downarrow , \, L\uparrow R\downarrow , \, C\uparrow R\downarrow}$ and ${L\downarrow C\uparrow , \, L\downarrow R\uparrow , \, C\downarrow R\uparrow}$ states cannot be straightforwardly analytically derived. However, setting ${\varepsilon_C=-(E_z^{L,R}-E_z^{C})/2}$ and expanding in a series in ${T_{ij}/|E_{C,\downarrow}-E_{R,\downarrow}|}$ leads to all states being decoupled, and the following time evolution operator $\tilde{U}=\text{diag}\left(\exp{\left(i(E_z^L-E_z^C)t/\hbar\right)},\exp{\left(i(E_z^L-E_z^C)t/\hbar\right)}\right)$ in the relevant basis of $\{|L,\downarrow; C\uparrow\rangle, |C,\uparrow;R,\downarrow\rangle\}$ (I used only $E_z^L$ from this point on because I assumed identical charge qubit quantum dots).

At the optimal $t=\tau$ the time evolution operator becomes

\begin{equation}
\mathcal{U}=
\begin{pmatrix}
0 && -1 &&  0 && 0\\
-1 &&0 && 0 && 0 \\
0 && 0 && e^{i\sqrt{2}\frac{\pi}{T}\left(E_z^C-E_z^L\right)} && 0\\
0 && 0 && 0 && e^{i\sqrt{2}\frac{\pi}{T}\left(E_z^C-E_z^L\right)} 
\end{pmatrix},
\end{equation}
	
 in the relevant basis of ${\{|L,\downarrow;C,\downarrow\rangle\,|C,\downarrow; R,\downarrow\rangle,\, |L,\downarrow; C\uparrow\rangle, |C,\uparrow;R,\downarrow\rangle\}}$. This is the CNOT gate between the spin and the charge qubit, multiplied by a single qubit operation on the spin qubit $\exp{\left(i(\phi\pm\pi)\sigma_z/2\right)}$, where $\phi=\sqrt{2}\frac{\pi}{T}\left(E_z^C-E_z^L\right)$. Single qubit gates alongside with the CNOT gate allow one to create any of the four Bell states \cite{Nielsen1}. 

\section{3) Fidelity of operations} I consider two electrons in a triple quantum dot, with the lowest orbital energy state included, two Zeeman sublevels in each dot, and states where both electrons are in $|\!\uparrow\rangle$ state excluded, 
yielding a total of 12 possible states. In contrast to a derivation in the previous section here all states are kept in the simulation in order to test the validity of the applied approximations in the derivation.

\begin{table}[h!]
\begin{tabular}{|c|c|c|c|c}
\multicolumn{2}{|c|}{initial state}&\multicolumn{2}{c|}{final state}\\
  \hline\hline
 $|1\rangle_s |0\rangle_c$&$|L,\downarrow;C,\downarrow\rangle$&$|C,\downarrow;R,\downarrow\rangle$&$e^{i(\phi\pm\pi)}|1\rangle_s |1\rangle_c$\\
 $|1\rangle_s |1\rangle_c$&$|C,\downarrow;R,\downarrow\rangle$&$|L,\downarrow;C,\downarrow\rangle$&$e^{i(\phi\pm\pi)}|1\rangle_s |0\rangle_c$\\
 $|0\rangle_s |0\rangle_c$&$|L,\downarrow;C,\uparrow\rangle$&$|L,\downarrow;C,\uparrow\rangle$&$|0\rangle_s |0\rangle_c$\\
 $|0\rangle_s |1\rangle_c$&$|C,\uparrow;R,\downarrow\rangle$&$|C,\uparrow;R,\downarrow\rangle$&$|0\rangle_s |1\rangle_c$
\end{tabular}
  \caption{The evolution of qubit states when the tunnel barriers are being subjected to a shuttling event with a duration $\tau$.}
  \label{Tab1}
\end{table}

In table Tab \ref{Tab1} I show the evolution of all possible states belonging to the qubit subspace, when the system is being subjected to a tunneling event of a duration $\tau$.
In the limit of tunneling smaller than Zeeman energy mismatch, the transitions ${|L,\downarrow;C\uparrow\rangle\longleftrightarrow|C,\uparrow;R,\downarrow\rangle}$ are not occurring due to the fact that the electron spin in the $|C,\uparrow\rangle$ state cannot tunnel to any other state due
to the energy mismatch with the energy of all other available $|\!\uparrow \rangle$ states.
Furthermore, the Coulomb penalization for charging a quantum dot with two electrons is much larger then the tunneling ${U\sim\text{ meV}\gg T}$, preventing the ${|L,\downarrow\rangle\leftrightarrow|R,\downarrow\rangle}$ transitions
via the central dot.

From Tab \ref{Tab1} we see that if the spin qubit is in the $|1\rangle_s$ state the state of the charge qubit is changed. Furthermore, if the spin qubit is in the $|0\rangle_s$ the state of the charge qubit remains unchanged. It should be noted that the 
spin qubit acquired a local phase of a $\phi\pm\pi$ in the spin up state.
Thus, the operation presented here is a two qubit CNOT operation times a single spin qubit $\phi\pm\pi$ phase gate (to the spin up state) where the spin qubit acts as a control qubit and the charge qubit acts as a target qubit.
The phase of $\exp{(i\phi)}$ is acquired due to the fact that when the electron spin is in the $|\!\uparrow\rangle$ state there is an energy mismatch leading to an accumulation of a local phase. The additional operation of $\exp{(\pm i\pi)}$ exists because the system is cycled half way through a cycle. This local phase can be corrected with a single qubit $Z$ gate on the spin qubit, and is rigorously analytically defined towards the end of the previous Section.

Fidelity of a quantum operation can be calculated as \cite{Pedersen1}

\begin{equation}\label{eq:fid}
\mathcal{F}=\frac{1}{n(n+1)}\big[{\rm Tr}(KK^{\dagger})+|{\rm Tr(K)}|^2\big].
\end{equation}
In the case of the CNOT gate, the dimensions of the Hilbert subspace are $n=4$ and, 
\begin{equation}
K={\rm CNOT}\,M_{q}\,U_s\,e^{-i \tilde{H}\tau/\hbar}\,M_{q}^{\dagger}.
\end{equation}
Here, ${\rm CNOT}$ is the matrix representation of an ideal CNOT gate, $M_{q}$ is a projection operator that projects out the $4$-dimensional qubit subspace from the $12$-dimensional Hilbert space of the Hamiltonian $\tilde{H}$, and $U_s=\exp{\left(-i(\phi\pm\pi)\sigma_z/2\right)}$ 
is the correcting gate accounting from cycling the acquired accumulated phase on the spin qubit. 

Combining the CNOT and the phase $U_s$ operation with single qubit Hadamard gates allows us to construct a SWAP gate [\onlinecite{Nielsen1}], and therefore embeds a general superposition of charge states into the spin degree of freedom. 
The SWAP gate fidelity is calculated by inserting

\begin{figure}[t!]
	\centering
	\includegraphics[width=0.4\textwidth]{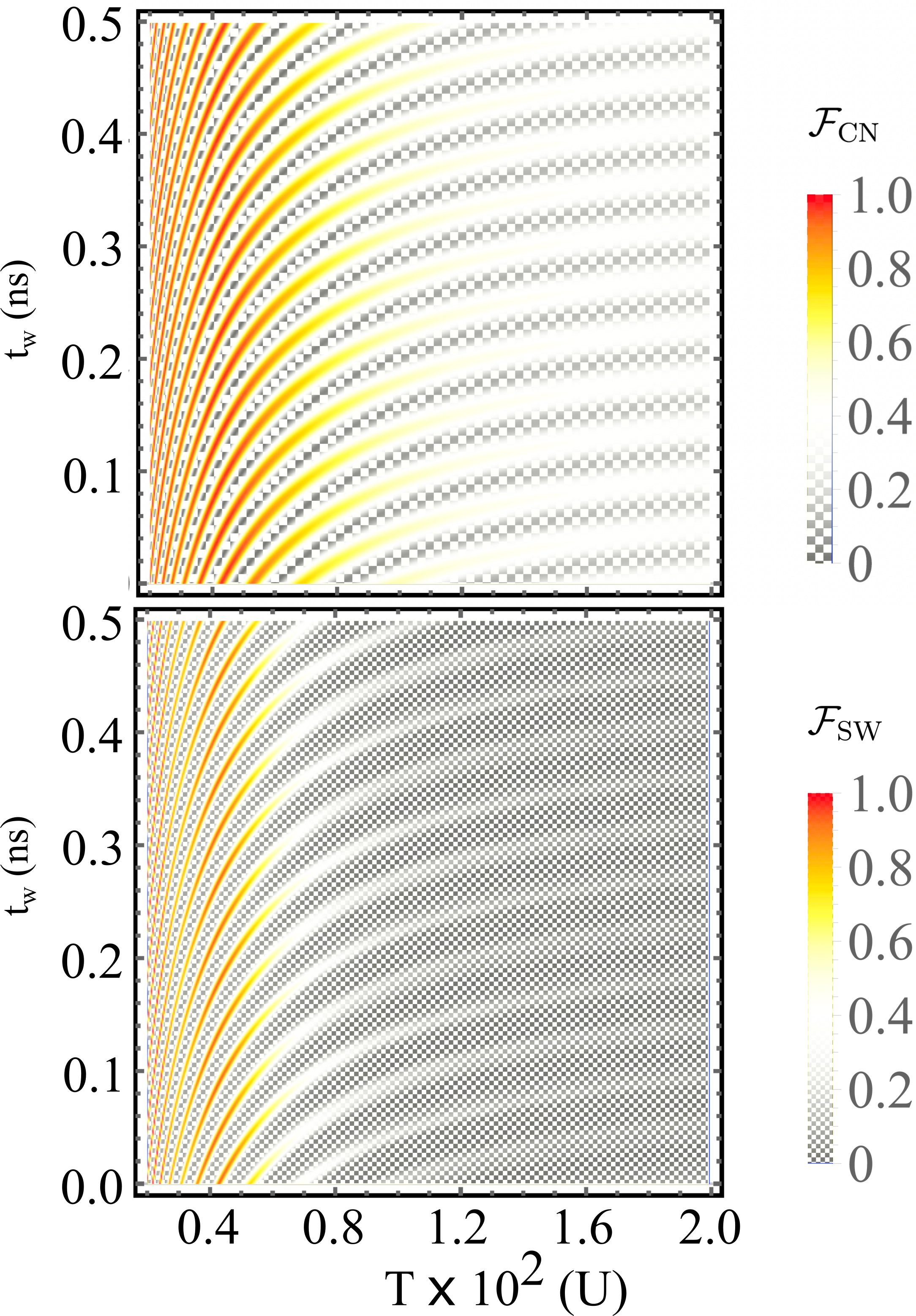}	
	\caption{(Color online) CNOT$\times R_{\pi}^{\rm s}$ and SWAP gate fidelities as a function of the tunneling $T$ and waiting time $t_w$. The parameters of the plot are the Coulomb repulsion of a doubly occupied dot $U=3\text{ meV}$, the Coulomb repulsion of a separated electrons $V =0.1\text{ meV}$, 
		the external magnetic field $B^z=-1\text{ T}$ and the magnetic field of the micromagnet $B_{\rm}=0.2\text{ T}$. A $g$-factor corresponding to Si quantum dots was utilized $g=2$. }
	\label{Res1}
\end{figure}

\begin{multline}\label{eq:Q}
K^{\rm SWAP}= {\rm SWAP} M_{q}U_se^{-i \tilde{H}\tau/\hbar}H_sH_cU_se^{-i\tilde{H}\tau/\hbar}H_sH_c\\ 
\times U_s e^{-i \tilde{H}\tau/\hbar}M_{q}^{\dagger},
\end{multline}
into Eq. (\ref{eq:fid}), for $n=4$. Here, ${\rm SWAP}$ is the matrix representation of an ideal SWAP gate, $M_{q}$ is a projection operator that projects out the 
$4$-dimensional qubit subspace from the $12$- space of the Hamiltonian $\tilde{H}$, and $H_s$ and $H_c$ are single qubit Hadamard gates of the spin and charge qubit respectively.

In an actual physical setting, the single spin Z-gates are performed by freely waiting for the spin to evolve in an external magnetic field for a certain time which we call the waiting time $t_w$, $U_Z=\exp{\left(iE_z^C t_w\sigma_z/2\hbar\right)}$. As the tunneling changes, a different local phase is acquired, requiring another value of the $Z$ rotation to correct it, consequently leading to a nontrivial dependence on the fidelity on the waiting time and tunneling see Fig. \ref{Res1}.  We see that both the CNOT and SWAP gate fidelities reduce with increasing the tunneling Fig. \ref{Res1}. When the tunneling is modified, this in turn changes the shuttling time $\tau$. In the limit of strong driving, the electron has enough energy to overpower the Coulomb penalization and can tunnel through the central quantum dot, although the central dot is occupied with an $|\!\uparrow\rangle$ electron. It should be noted that in our scheme the fidelity of the SWAP gate is always lower then the fidelity of the 
CNOT gate due to the fact that the SWAP gate is achieved as a combination of three CNOT operations, and is therefore more susceptible to leakage to logical subspaces not comprising our qubit states. 

As the charge qubit is usually more susceptible to noise than the spin qubit, a crucial requirement for a successful implementation of a SWAP operation is that the state of the charge qubit remains coherent thought the duration of the sequence of pulses conducted to achieve the SWAP gate $T_{2,c}^{*} > 3 \tau+2 \max{( T_{H_{s}},T_{H_{c}})}$ . In typical gate defined quantum dots, the charge qubit coherence time is $T_{2,c}^* \sim 10 \text{ ns} $, the spin qubit coherence time is $T_{2,s}^{*} \leq 200\text{ $\mu$s} $, and single qubit spin and charge Hadamard gate times, as low as $T_{H_{s}} \sim 1\text{ ns}$ \cite{froning2021ultrafast}, $T_{H_{c}} \sim 1 \text{ ns}$, while the parameter $\tau\le0.1\text{ ns}$. 
However, the charge qubit coherence time of an isolated DQD charge qubit was measured to be $T_{2,c}^{*}=220\text { ns}$ \cite{Gorman1} with a full rotation on the Bloch sphere being achieved for $\sim 50\text{ ns}$, so these systems represent a viable candidate for using the spin degree of freedom of a long-lived memory of the charge state. In a more recent study \cite{PhysRevLett.122.206802}, charge qubits coupled to microwave resonators achieve a coherence time of $T_{2,c}\sim 50\text{ ns}$ with qubit operation times of $5\text{ ns}$, representing another viable system for the realization of the scheme.

\section{4) Conclusion} To conclude, I have demonstrated an entangling scheme for a spin and charge degree of freedom in semiconductor quantum dots. The combination of CNOT gates allows us to construct a SWAP gate and therefore, use the spin degree of freedom as a long lived memory of the charge state. Further research in this direction will focus on investigating the decoherence effects originating from charge noise and nuclear spins.

\section{5) Acknowledgments} 
I thank Maximilian Russ, Guido Burkard, Heng Wang and Thierry Ferrus for useful suggestions

\bibliographystyle{apsrev}
\bibliography{PaperV_1}
\end{document}